\documentclass[aps,prl,notitlepage]{revtex4-1}
\usepackage{bm}
\usepackage{graphicx,epsfig,color}

\usepackage{amssymb,amsfonts,amsmath}

\bibpunct{}{}{,}{s}{}{,}

\newcommand{\bs}{\mathbf {s}}
\newcommand{\br}{\mathbf {r}}

\newcommand{\etal}{{\it et al.}~}

\begin{document}

\title{A note on the propagation of quantized vortex rings through a quantum turbulence tangle: Energy transport or energy dissipation?}
\author{Jason Laurie$^1$, Andrew W. Baggaley$^2$}

\affiliation{$^1$Department of Physics of Complex Systems, Weizmann Institute of Science, 234 Herzl Street, Rehovot 76100, Israel\\ 
$^2$School of Mathematics and Statistics, University of Glasgow, 15 University Gardens,
Glasgow, G12 8QW, UK\\
}

\date{\today}


\begin{abstract}
We investigate quantum vortex ring dynamics at scales smaller than the inter-vortex spacing in quantum turbulence. Through geometrical arguments and high resolution numerical simulations we examine the validity of simple estimates of the mean free path and the structure of vortex rings post-reconnection. We find that a large proportion of vortex rings remain coherent objects where approximately $75\%$ of their energy is preserved.  This leads us to consider the effectiveness of energy transport in turbulent tangles. Moreover, we show that in low density tangles, appropriate for the ultra-quantum regime, ring emission cannot be ruled out as an important mechanism for energy dissipation. However at higher vortex line densities, typically associated with the quasi-classical regime, loop emission is expected to make a negligible contribution to energy dissipation, even allowing for the fact that our work shows rings can survive multiple reconnection events. Hence the Kelvin wave cascade seems the most plausible mechanism leading to energy dissipation.
\end{abstract}

\maketitle

\section{Introduction}

The study of vortex rings is an old one, dating back at least to the 1850s~\cite{rogers_formation_1858}; despite a long history they still provide a very active area of fluid dynamics research~\cite{shariff_vortex_1992,kleckner_creation_2013}. One system which provides the ideal playground to investigate the dynamics of vortex rings are those of quantum fluids such as superfluid helium and atomic Bose-Einstein condensates. The reasons are two-fold, firstly quantized vortices are stable topological defects, with a fixed circulation and secondly, particularly in superfluid helium, the vortex core size is typically orders of magnitude smaller than the radius of the vortex ring~\cite{donnelly_quantized_1991}. This is in stark contrast to vortex rings in classical fluids where both viscous dissipation and instabilities due to their (relatively) thick cores severely constrain the lifetime of even an isolated ring. Another active field of research in quantum fluids is that of quantum turbulence~\cite{barenghi_introduction_2014}, where vortex rings offer an ideal experimental tool to both generate~\cite{bradley_emission_2005,walmsley_quantum_2008} and probe~\cite{walmsley_quantum_2008} the complex fluid motion. Moreover, recently a great deal of attention has focused on the role that vortex rings may play in the decay of quantum turbulence~\cite{kondaurova_numerical_2012,nemirovskii_decay_2013}.

Svistunov~\cite{svistunov_superfluid_1995} suggested that a reconnection between two almost anti-parallel vortices can lead to the emission of a series of vortex rings through self-reconnections of the vortex lines.  Indeed, subsequent numerical simulations performed by both Kerr~\cite{kerr_vortex_2011} and Kursa~\etal~\cite{kursa_cascade_2011} have confirmed this scenario. With this in mind, it is plausible to think~\cite{nemirovskii_decay_2013} that this mechanism may provide an alternative route for energy dissipation other than a Kelvin wave cascade~\cite{kozik_kelvin-wave_2004,lvov_weak_2010,baggaley_kelvin-wave_2014}. So far, numerical evidence suggests that reconnections between anti-parallel vortex reconnections dominate in tangles that contain no large-scale vortex bundles~\cite{baggaley_thermally_2012}, albeit with angles far higher than the critical angle derived by Svistunov~\cite{svistunov_superfluid_1995}. Further numerical studies by Kondaurova and Nemirovskii~\cite{kondaurova_numerical_2012} have also highlighted the possible importance of vortex ring emission in the decay of the random, unstructured, Vinen tangles also known as the ultra-quantum regime. 

In contrast, for the case of the much studied quasi-classical Kolmogorov picture of homogeneous and isotropic quantum turbulence (at $0$~K), vortex ring emission is not expected to be as important~\cite{kozik_kolmogorov_2008}. Here energy is injected into the fluid at large scales (scales greater than the typical vortex separation) and locally cascades in a Richardson like manner through scales by the breakup of polarized vortex bundles~\cite{baggaley_importance_2012}. This process is consistent until energy reaches the typical scale of the inter-vortex spacing $\ell$, where the notion of a vortex bundle breaks down. At this scale, energy still cannot be dissipated, and must be transfered to even smaller scales.  This superfluid cross-over mechanism is still somewhat unknown~\cite{lvov_bottleneck_2007,kozik_theory_2009}, but vortex reconnections are thought to play an essential role. The general consensus is that energy eventually arises in the form of small-scale helical waves, known as Kelvin waves, that propagate along quantized vortex lines.  These waves weakly interact to excite even smaller scale waves until reaching frequencies capable of exciting phonons that dissipate energy as heat. 

It is typical (and appealing) to apply simple geometrical arguments to investigate the importance of vortex ring emission in the decay of quantum turbulence, see for example arguments concerning the `opaqueness' of a tangle in~\cite{kursa_cascade_2011}. For instance, if we take a tangle of quantized vortices of total length $\Lambda$, confined within a volume $V$ then we define the vortex line density as $L=\Lambda/V$. If a single vortex ring is created (through a vortex reconnection for example) with radius $r$, then the probability per unit time that the ring will meet, and hence reconnect, with a vortex line is $2rLv_{\rm ring}$, where $v_{\rm ring}$ is the velocity of the vortex ring. If one considers a circular vortex ring without Kelvin wave distortions and neglects the influence of other vortices by only considering the self-induced motion of the vortex ring, which is perhaps justified for radii smaller than the mean inter-vortex spacing $\ell$ and larger than the vortex core radius, then the speed of a quantized vortex ring is given by
\begin{equation}\label{eq:ring_speed}
v_{\rm ring}=\frac{\Gamma}{4 \pi r} \left[ \ln{(8r/a)}-\alpha \right],
\end{equation}
where $a$ is the core radius ($a\approx 10^{-8}\,$cm for superfluid $^4$He), 
$\Gamma$ is the quantum of circulation ($\Gamma=9.97\times 10^{-4}\,{\rm cm}/{\rm s}^2$ for superfluid $^4$He) and $\alpha$ is an order one constant that depends on the assumed vortex core structure~\cite{donnelly_quantized_1991}.

From Eq.~(\ref{eq:ring_speed}), one can estimate the mean free path $\left\langle d\right\rangle$ of a vortex ring of radius $r$ in a vortex tangle of vortex line density $L$ as
\begin{equation}\label{eq:MFP}
\left\langle d\right\rangle=\frac{1}{2rL}.
\end{equation}

The mean free path estimate has been used to both interpret experimental results \cite{walmsley_reconnections_2014} and determine theoretically the importance of ring emission to the decay of quantum turbulence \cite{kursa_cascade_2011}.  However, as far as the current authors are aware, the mean free path estimate~(\ref{eq:MFP}) has never been under scrutiny itself.  Indeed, there are a number of reasons one may question the validity of Eq.~(\ref{eq:MFP}). Firstly, nonlocal interactions between the vortex ring and the vortex tangle may substantially change the dynamics of the ring; for example, an approaching ring could be deflected around vortex lines before reconnection. Secondly, the vortex ring could reconnect with a part of the tangle, but leave a considerable part of the ring free to propagate further. Or thirdly, even if the vortex ring is fully absorbed back into the tangle, one could imagine that large amplitude Kelvin waves would be generated triggering self-reconnections and further vortex ring emissions~\cite{salman_breathers_2013}.

Our main motivation is to study the mean free path estimate of Eq.~(\ref{eq:MFP}) in detail and test it numerically. In addition, we also examine some of the possible scenarios outlined in the previous paragraph to delve deeper into quantum vortex ring dynamics.

The structure of the paper is as follows: In section~\ref{sec:num} we detail the numerical strategy for all the simulations, followed by the examination of the validity of the mean free path estimate in section~\ref{sec:vortexring}. In section~\ref{sec:ring_line} we consider an idealised ring-line reconnection and derive predictions for the post-reconnection structure of a ring based on a simple geometric argument, before numerically studying ring generation in a turbulent tangle in section~\ref{sec:ringscale}.  Finally, we bring all the results of the previous sections together to discuss energy transport by vortex rings in typical experimental setups in section~\ref{sec:energytrans} before concluding in section~\ref{sec:con}.

\section{Numerical techniques}
\label{sec:num}

The vortex filament method of Schwarz~\cite{schwarz_three-dimensional_1985} models quantum turbulence by approximating quantized vortex lines as one-dimensional space curves ${\bs}={\bs}(\xi,t)$ where $\xi$ is arc length and $t$ is time. At zero temperature, where mutual friction effects are absent and in the regime where there is no external superfluid flow, the dynamics of the quantized vortex lines are determined by the Biot-Savart Law
\begin{equation}
\frac{d\bs}{dt}=\frac{\Gamma}{4 \pi} \oint_{\cal L} \frac{(\br-\bs) }
{\left| \br - \bs \right|^3}
\times {\bf d}\br.
\label{eq:BS}
\end{equation}
Here, we use parameters that correspond to pure superfluid $^4$He: circulation 
$\Gamma=9.97 \times 10^{-4}~\rm cm^2/s$ and a vortex core radius
$a= 1\times 10^{-8}~\rm cm$, but our results can be generalized to 
turbulence in low temperature $^3$He-B. 

The line integral in Eq.~(\ref{eq:BS}) extends over the entire vortex configuration $\cal L$,
which is discretized into a large number of points $\bs_i$ where $i=1,\cdots, N$.  The Biot-Savart law~(\ref{eq:BS}) contains a singularity when $\br=\bs$, which we regularize in a standard way 
by considering the local and non-local contributions to the integral separately. Consequently, if we denote  the position of the $i^{\rm th}$ discretization point as $\bs_i$ along the vortex line, then Eq.~(\ref{eq:BS}) becomes
\begin{equation}
\frac{d\bs_i}{dt}=
\frac{\Gamma}{4\pi} \ln \left(\frac{\sqrt{\ell_i \ell_{i+1}}}{a_0}\right)\bs_i' \times \bs_i'' 
+\frac{\Gamma}{4 \pi} \oint_{\cal L'} \frac{(\br-\bs_i) }
{\left| \br - \bs_i \right|^3}
\times {\bf d}\br.
\label{eq:BS_sing}
\end{equation}
Here $\ell_i$ and $\ell_{i+1}$ are the arc lengths of the curve 
between points $\bs_{i-1}$ and $\bs_i$ and between $\bs_i$ and $\bs_{i+1}$ respectively, 
and $\cal L'$ represents the remaining (nonlocal) vortex tangle.

The precise details of the techniques on how we discretize the vortex lines
into a variable number of points $\bs_i$ where, $i=1,\ldots, N$ held at minimum separation distance of $\Delta \xi/2$
are described in~\cite{baggaley_spectrum_2011}, whilst information on how we implement the artificial vortex reconnections is found in~\cite{baggaley_sensitivity_2012}. All spatial derivatives are calculated using a fourth-order finite difference scheme, and time-stepping is achieved with a third-order Runge-Kutta method. 

If one neglects the nonlocal integral of Eq.~(\ref{eq:BS_sing}), then the velocity of a vortex point is induced solely by the neighbouring vortex line segments and the model is known as the local induction approximation (LIA). LIA has been used in several works~\cite{schwarz_three-dimensional_1988,kondaurova_numerical_2008} due to its vastly superior computational efficiency to the full Biot-Savart law~(\ref{eq:BS}). That said, it have been shown~\cite{adachi_numerical_2011} to fail in reproducing the correct physics in many situations.

\section{Vortex ring propagation through a tangle}
\label{sec:vortexring}

To begin, we wish to investigate the validity of the mean free path estimate of Eq~(\ref{eq:MFP}). One may expect that the distribution of the vortex ring propagation distance before reconnection to be exponentially distributed if nonlocal effects are small. With this in mind, given the prior estimate of the mean free path of the vortex ring Eq.~(\ref{eq:MFP}), and assuming an exponential distribution, the probability of the vortex ring to propagate a distance $W$ within the tangle is given by
\begin{equation}\label{eq:prob_exp}
P(d=W) = 2rL\exp\left(-2rLW\right),
\end{equation}
where we have used the formula for the mean free path~(\ref{eq:MFP}).  A more natural and convenient reformulation of Eq.~(\ref{eq:prob_exp}) is in the cumulative probability distribution of observing a vortex ring of radius $r$ successfully navigating a tangle of width $W$ and density $L$ without reconnecting.  This probability can be found from Eq.~(\ref{eq:prob_exp}) by
\begin{equation}\label{eq:prob_cum}
P(d>W) = \int^{\infty}_W \, P(d')\, {\rm d}d' = \exp\left(-2rLW\right).
\end{equation}
Note, that the cumulative probability~(\ref{eq:prob_cum}) neglects the temporal dynamics of the vortex tangle as the vortex ring propagates. If fact, if we take into account this diffusion the result will not change as the spread of the tangle is compensated by the reduced vortex line density.

\begin{figure}[htbp]
	\centering
	\includegraphics[width = 0.49\textwidth]{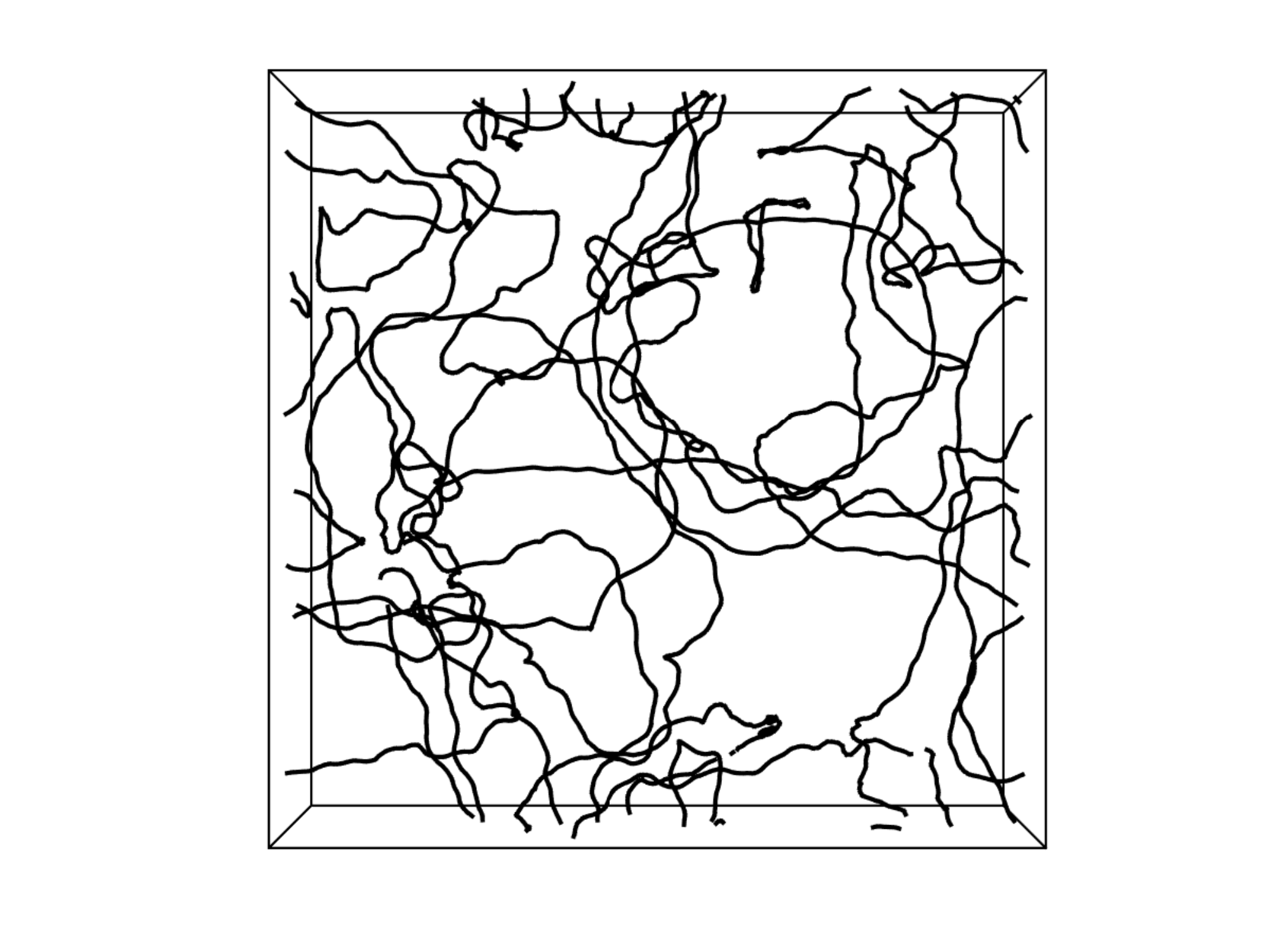}
	\hfill
	\includegraphics[width = 0.49\textwidth]{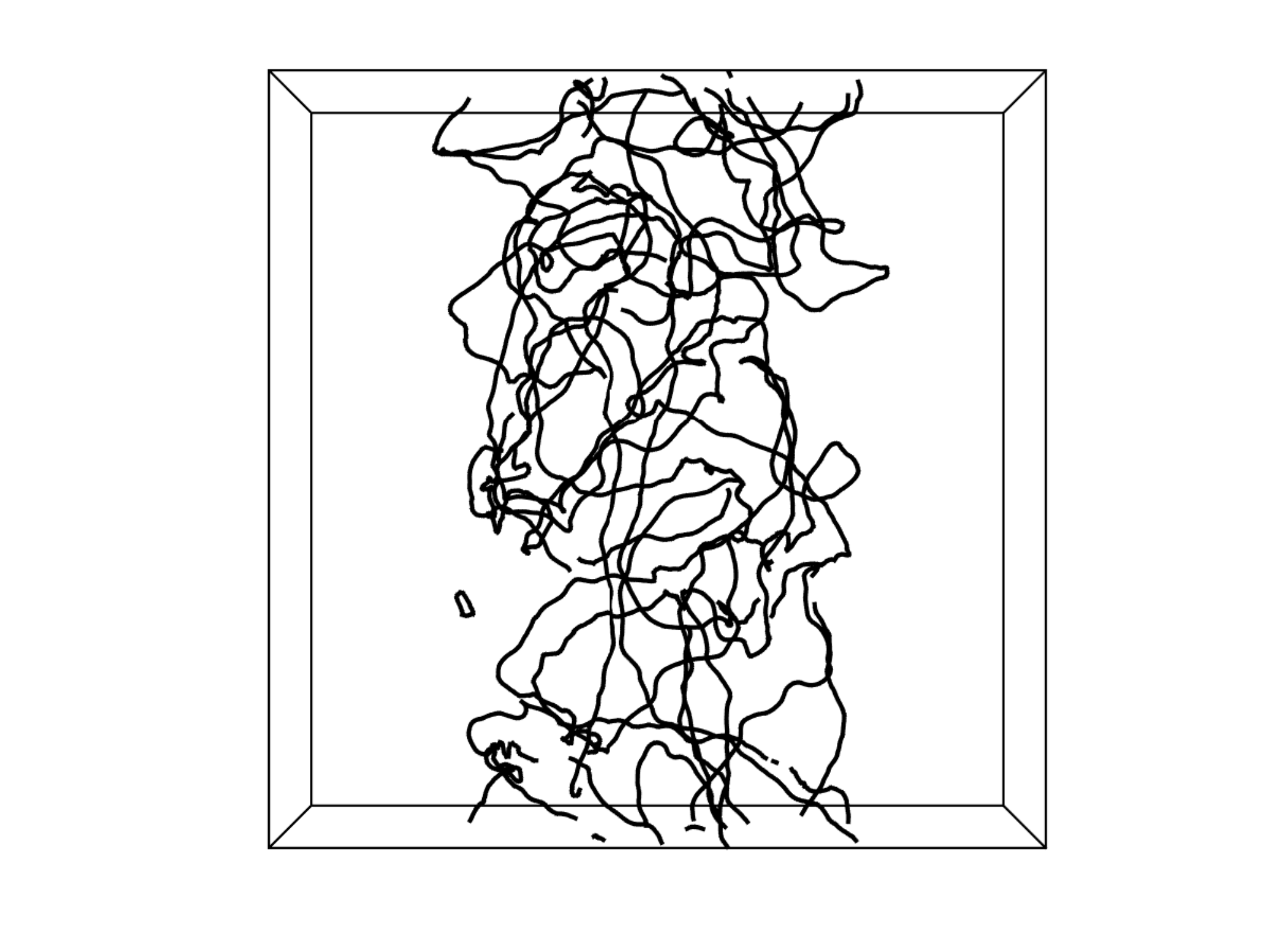}
	\caption{Image of the vortex tangle created at the center of the numerical box across the $xy$-plane. View of the tangle along the $z$-plane (left) and along the $x$-plane (right). }
	\label{fig:tangle}
\end{figure}

To check the hypothesis that the free path is exponentially distributed we perform a numerical experiment to measure the cumulative probability density function~(\ref{eq:prob_cum}).  We begin by creating a random tangle at the centre of the numerical box. To do this we initialize the numerical simulation with a large number of randomly oriented vortex rings, uniformly distributed in a small strip in the $xy$-plane of the numerical box around the origin with a width in $z$ is $1\times 10^{-1}~\rm cm$. The system is time evolved using the Biot-Savart law for a sufficient period until the mean curvature is saturated and measures of the relative (by length) anisotropy of the tangle are statistically equal.  Fig.~\ref{fig:tangle} displays the resulting vortex tangle. The tangle contains a total vortex line length of $\Lambda=35.4~\rm cm$ inside a volume of $V=0.6~{\rm cm} \times 1~{\rm cm} \times 1~{\rm cm}$, giving a vortex line density of $L= \Lambda/V = 5.88\times 10^1~{\rm cm}^{-2}$.  The width of the tangle in the $z$-direction is $W \simeq 0.6~{\rm cm}$.

In order to compute the cumulative probability~(\ref{eq:prob_cum}), we perform a series of simulations using both the Biot-Savart law and the LIA, where we propagate vortex rings of three different radii across the vortex tangle in the $z$-direction. In total we perform a total of $800$ simulations for four different radii of vortex rings $r=9.0\times 10^{-3}$, $1.8\times 10^{-2}$, $3.6\times 10^{-2}$, and $5.4\times 10^{-2}~\rm cm$. The vortex tangle configuration displayed in Fig.~\ref{fig:tangle} is used as the initial condition with the injection of a single vortex ring of radius $r$, randomly positioned in the $xy$-plane at the edge of the box $z=-{\cal D}/2$, oriented such that the vortex ring propagates towards the vortex tangle. Each ring is tracked and if the initial ring reaches the opposite end of the box at $x={\cal D}/2$ without undergoing a vortex reconnection then the ring is deemed to have propagated successfully through the vortex tangle.  The spatial resolution used for these numerical runs is $\Delta \xi=5 \times 10^{-3}~\rm cm$ with a timestep of $\Delta t=1\times 10^{-4}~\rm s$.

\begin{figure}[htbp]
	\centering
	\includegraphics[width = 0.6\textwidth]{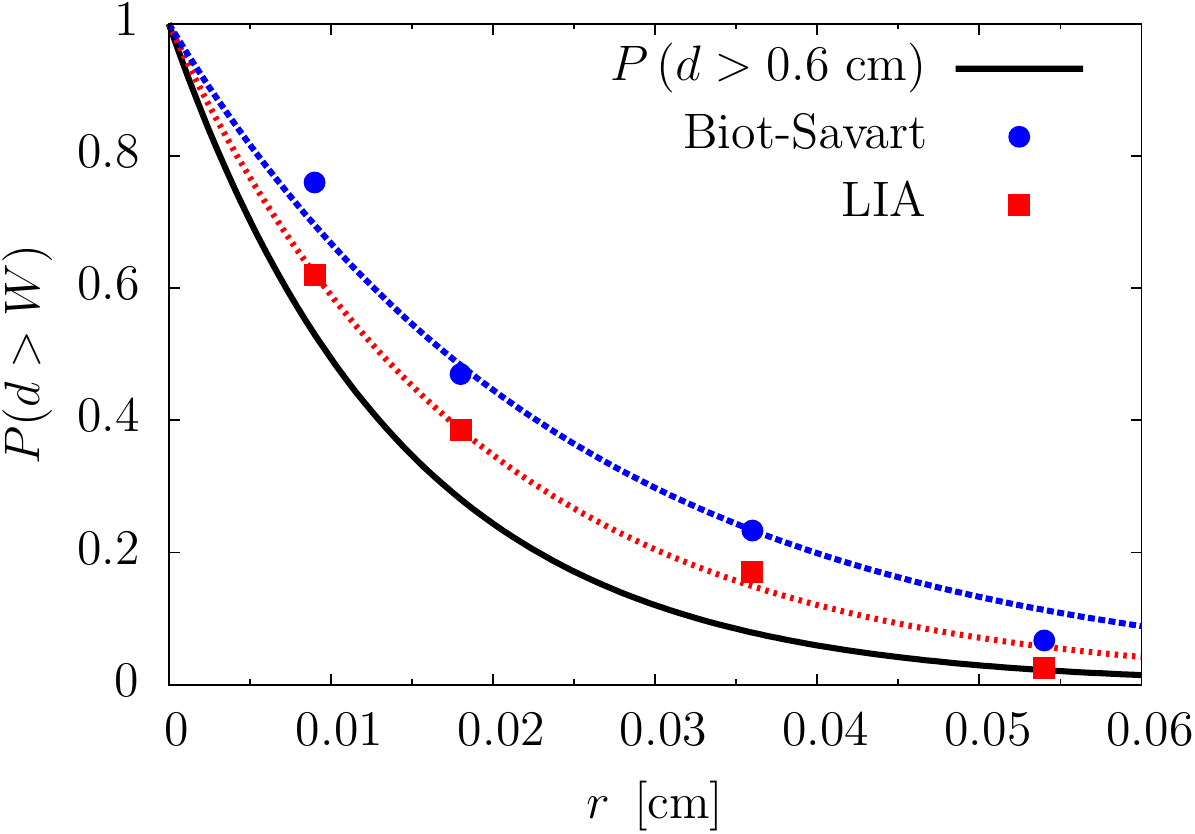}
	\caption{Cumulative probability of a vortex ring of radius $r$ transversing a random vortex tangle with initial width $dW=0.6~{\rm cm}$ without experiencing a vortex reconnection.  The numerical data for the Biot-Savart law (blue circles) and LIA (red squares) are plotted for four vortex ring radii of $r=9\times 10^{-3}$, $1.8\times 10^{-2}$, $3.6\times 10^{-2}$ and $5.4\times 10^{-2}~\rm cm$.  The solid black curve is the theoretical prediction Eq.~(\ref{eq:prob_cum}) which underestimates the numerical data. The blue dashed and red dotted curves are exponential fits $P=\exp(-crLW)$ with $c=1.14$ and $1.50$ to the Biot-Savart and LIA data respectively.}
	\label{fig:prob_survive}
\end{figure}

Figure~\ref{fig:prob_survive} shows the numerically obtained probabilities from our simulations using the Biot-Savart law and the LIA, and compares to the theoretical prediction of Eq.~(\ref{eq:prob_cum}) derived based on exponentially distributed reconnections and the mean free path estimate~\ref{eq:MFP}. We observe that both LIA and the Biot-Savart simulations leads to probabilities higher than those of the theoretical estimate. On a general basis we expect that LIA would yield probabilities more in line with the theory as nonlocal interactions between the ring and tangle are neglected. Although LIA produces probabilities closer to the theory there is still some discrepancy. The blue dashed and red dotted curves indicate numerical fits by least squares with an exponential distribution for the both the Biot-Savart and LIA data respectively. Interestingly, the LIA data fits the exponential distribution incredibly well, with a fitting that implies (to two decimal places) a mean free path of $\langle d\rangle=2/3rL$, $33\%$ larger than Eq.~(\ref{eq:MFP}). It is unclear if this exact prefactor is mere coincidence or if this indicates that the mean free path is underestimated. Furthermore, Fig.~\ref{fig:prob_survive} shows that the Biot-Savart model leads to an increase in success of transversing the vortex tangle without reconnection with an exponential distribution with mean $\langle d\rangle=1/1.14rL$.  Consequently, we can surmise that nonlocal effects must be responsible for an increased avoidance of the vortex ring to reconnect with the tangle.  This implies that reconnections between individual vortex rings and vortex tangles may be less common than previously thought. 

\begin{figure}[htbp]
	\centering
	\includegraphics[width = 0.6\textwidth]{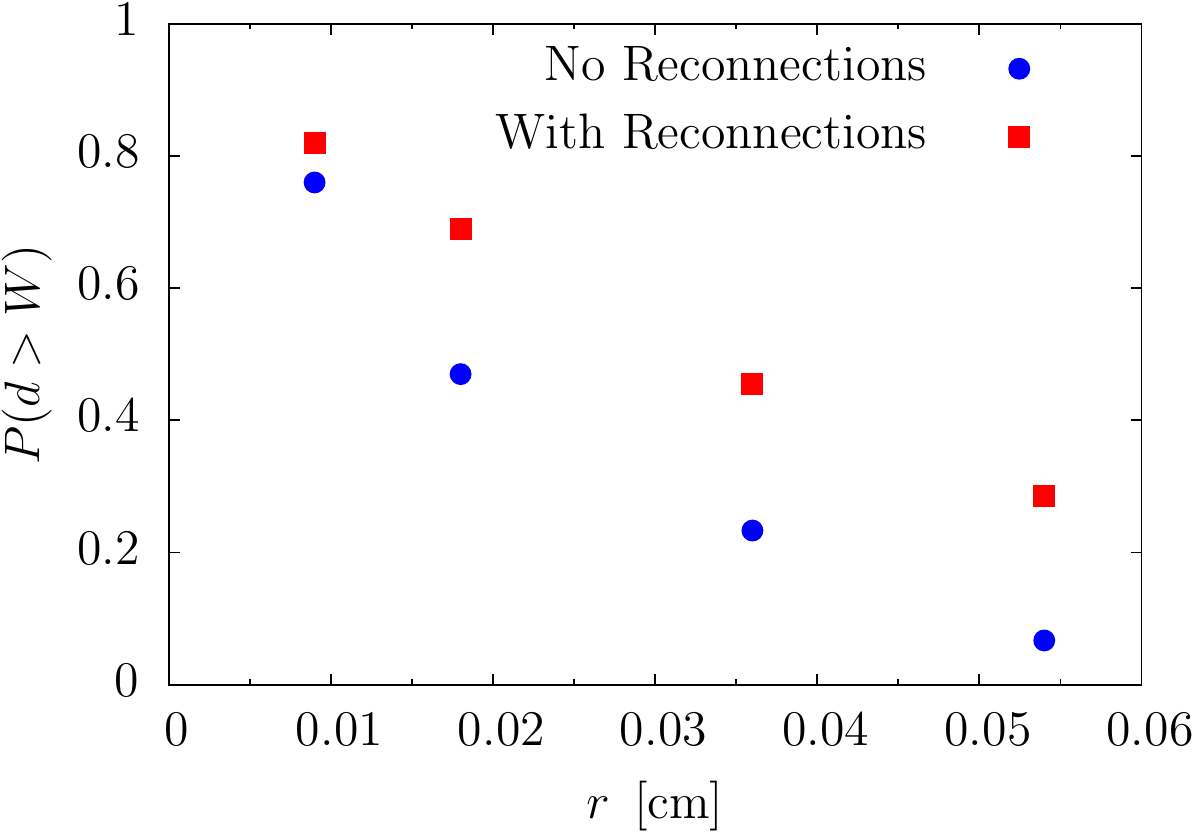}
	\caption{A comparison between the Biot-Savart data from Fig.~\ref{fig:prob_survive} (blue circles) with numerical data that include rings which reconnect with the tangle, but continue to propagate as coherent objects (red squares).}
	\label{fig:prob_survive_recon}
\end{figure}

Finally, we measure the total number of rings that propagate through the tangle, despite undergoing one, or perhaps multiple reconnections; this data is plotted in Fig.~\ref{fig:prob_survive_recon}.  We observe that there is still significant probability of a vortex ring being completely absorbed. However, numerical constraints limit the smallest ring radius we can resolve, and so it is very likely that, in reality, many more rings would propagate through the tangle. That being said, the fact that vortex rings can emerge as coherent objects after a reconnection means that significantly more energy can propagate through the tangle than one may naively expect based on the mean free path.

\section{Vortex ring-line reconnection \label{sec:ring_line}}

After considering the probability of a vortex ring reconnecting inside a vortex tangle, we now consider the reconnection behaviour between a vortex ring and vortex line. In this section, we will outline a simple geometric argument to estimate the post-reconnection structure of a vortex ring and compare to numerical data.

\begin{figure}[htbp]
	\centering
	\includegraphics[width = 0.6\textwidth]{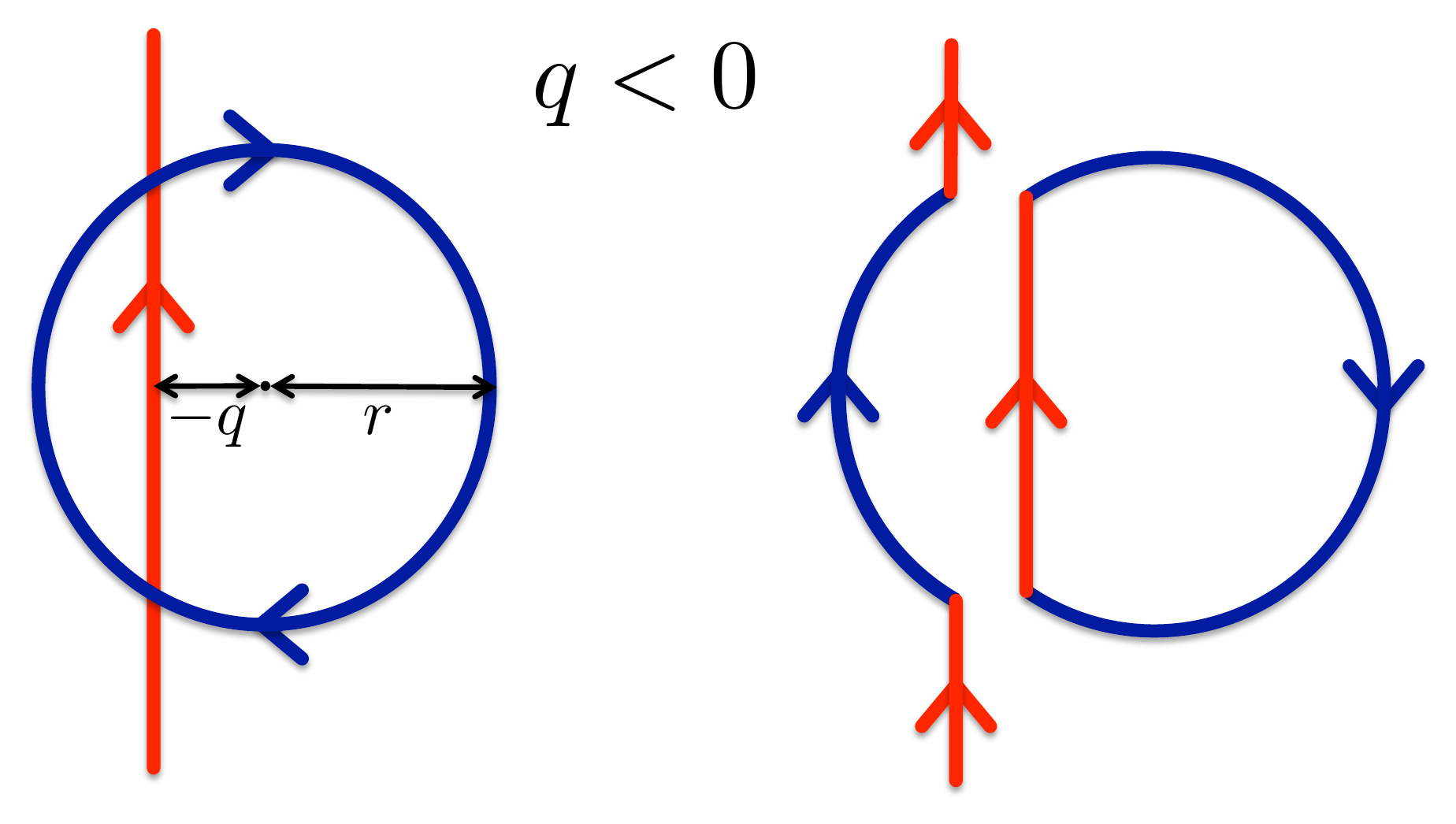}
	\includegraphics[width = 0.6\textwidth]{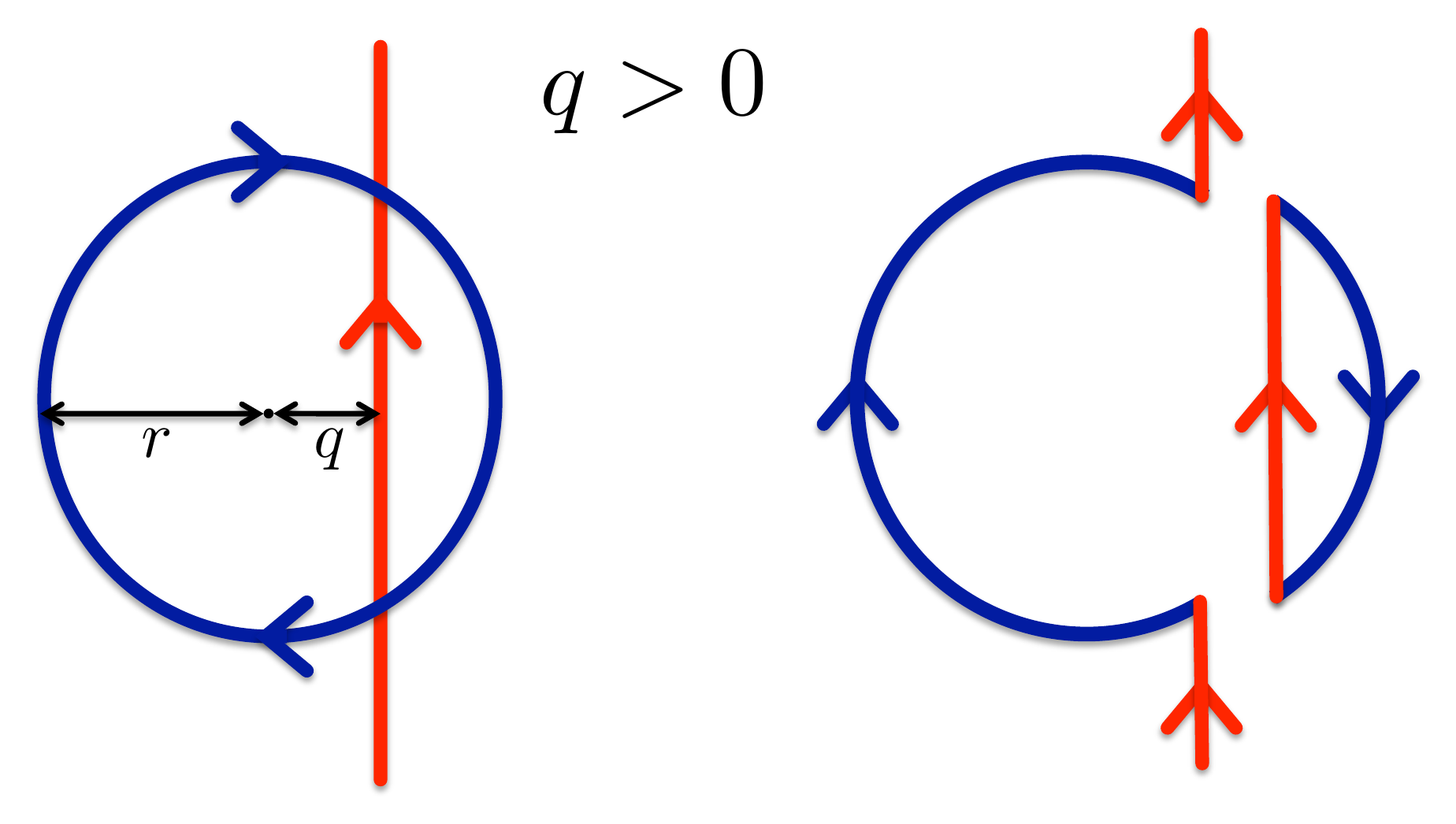}
	\caption{Schematic to show head on reconnection between a vortex ring (blue) and a vortex line (red) depends on the orientation.}
	\label{fig:ideal_recon}
\end{figure}

Consider a single vortex ring of radius $r$ propagating towards a vortex line, as depicted in Fig.~\ref{fig:ideal_recon}. We define the impact factor $q$ as the distance between the centre of the vortex ring and the axis of the vortex line. In such a setup the ring will propagate towards the vortex line and reconnect.  By assuming an ideal reconnection, as depicted in Fig.~\ref{fig:ideal_recon}, then we can compute the post-reconnection circumference $C$ of the surviving vortex ring as
\begin{equation}
C(q)=2\left(r^2-q^2\right)^{1/2}+2r\arccos\left(\frac{q}{r}\right),
\end{equation}
where the impact factor $q\in[-r,r]$.

If we further assume that the distribution of impact factors are uniform, as one may expect in a isotropic tangle, then we can calculated the expected value of the post-reconnection vortex ring circumference by integrating $C(q)$ over $q\in[-r,r]$ and dividing by $2r$:
\begin{equation}\label{eq:Cpredict}
\left\langle C\right\rangle=\frac{1}{2r}\int_{-r}^r \,C(q)\,{\rm d}q=\frac{3\pi r}{2}.
\end{equation}
Subsequently, we can compute the expected radius of the vortex ring after reconnection by dividing the average circumference (\ref{eq:Cpredict}) by $2\pi$:
\begin{equation}\label{eq:rpredic}
  \left\langle r_{\rm post}\right\rangle = \frac{\left\langle C\right\rangle}{2\pi} = \frac{3r}{4}.
\end{equation}

We remark that this geometric picture can provide a plausible explanation for an interesting feature noticed in the recent experimental study of Walmsley~\etal~\cite{walmsley_reconnections_2014}, where they measured the time of flight of charged vortex rings propagating through a quantized vortex tangle.  They discovered that some vortex rings were found to propagate through the system far faster than one would expect based on their initial radius. The explanation of Walmsley \etal was that reconnections could produce quantized vortex rings with a very small radius, orders of magnitude smaller than the initial injected ring, resulting in faster velocities as indicated by Eq.~(\ref{eq:ring_speed}). In our simple geometric picture, such tiny vortex rings are created when the impact factor is large $q\simeq r$.

\begin{figure}[htbp]
	\centering
	\includegraphics[width = 0.3\textwidth]{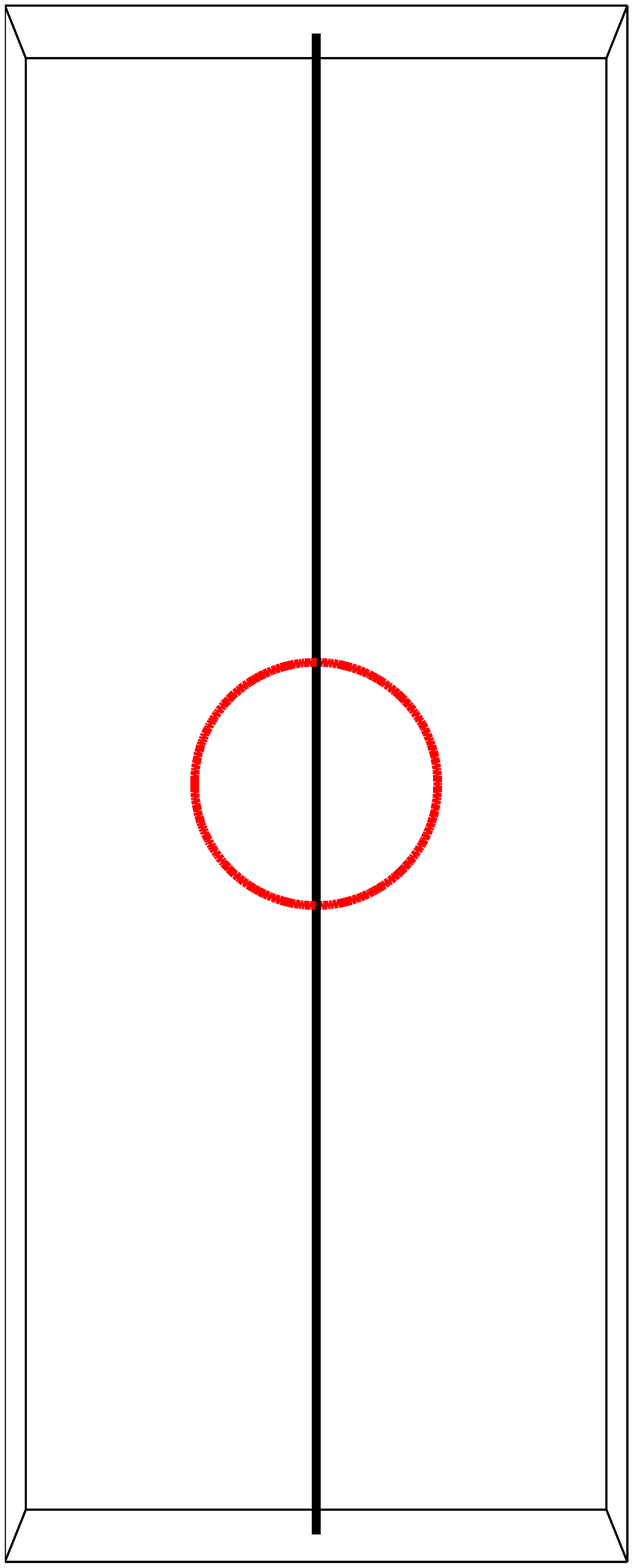}
	\includegraphics[width = 0.3\textwidth]{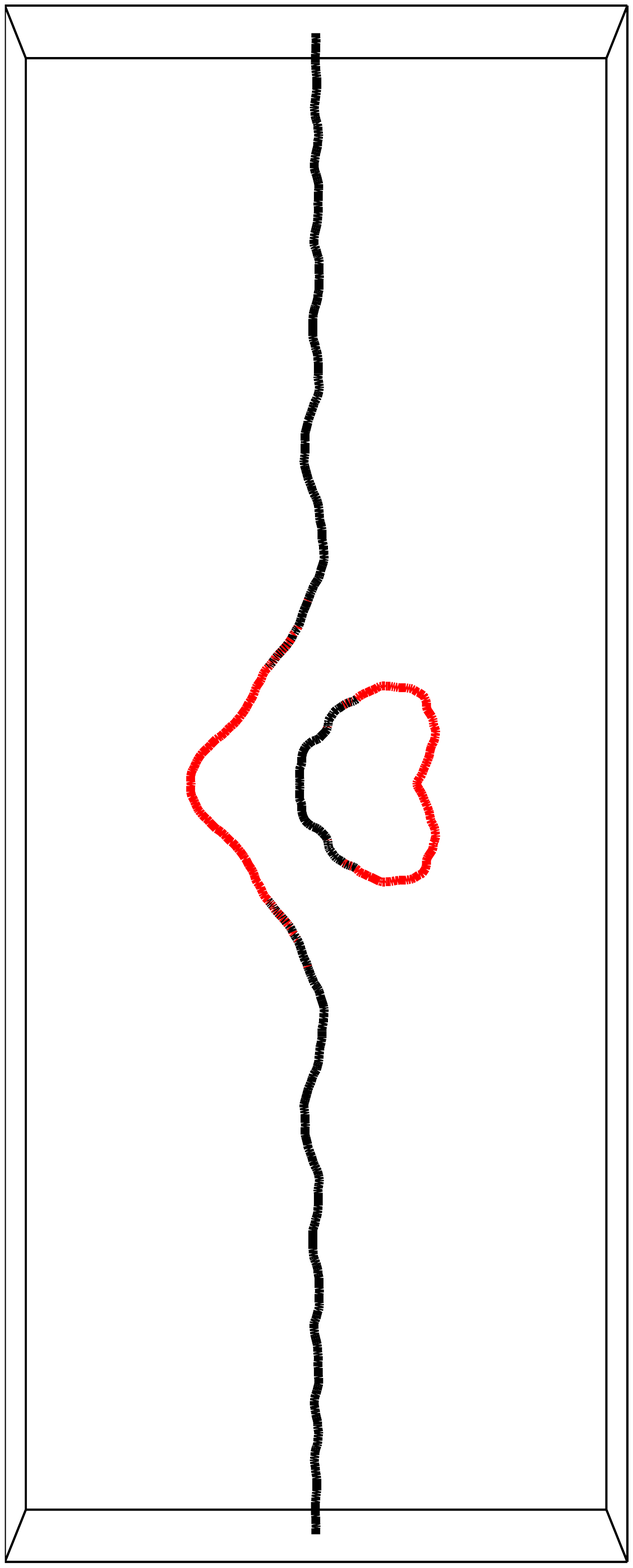}
	\caption{Initial (left) and final (right) state of vortex-line reconnection between a vortex ring (red) and vortex line (black) with $q=0$. Notice the exchange of vortex line segments between the ring and the line post-reconnection.}
	\label{fig:vortex-line-recon}
\end{figure}

To test our prediction (\ref{eq:Cpredict}), and hence Eq.~(\ref{eq:rpredic}), we perform a set of numerical simulations of vortex-line reconnections across a range of impact factors $q\in[-r,r]$. Numerically, we take a point separation of $\Delta \xi=1 \times 10^{-3}~\rm cm$ and use a timestep of $\Delta t=2.5\times 10^{-5}~\rm s$. The initial vortex ring has a radius of
$r=1/4\pi \simeq 8.0\times 10^{-2}~\rm cm$ (initially discretized into $500$ vortex segments) with a straight vortex line located in the centre of the box.  We vary the initial position of the vortex ring in order to cover a range of impact parameters $q$. After reconnection we allow the vortex ring to propagate until it reaches the edge of the numerical domain before computing its post-reconnection circumference. Fig.~\ref{fig:vortex-line-recon} shows the initial and final states of one of the simulations where $q=0$.

\begin{figure}[htbp]
	\centering
	\includegraphics[width = 0.55\textwidth]{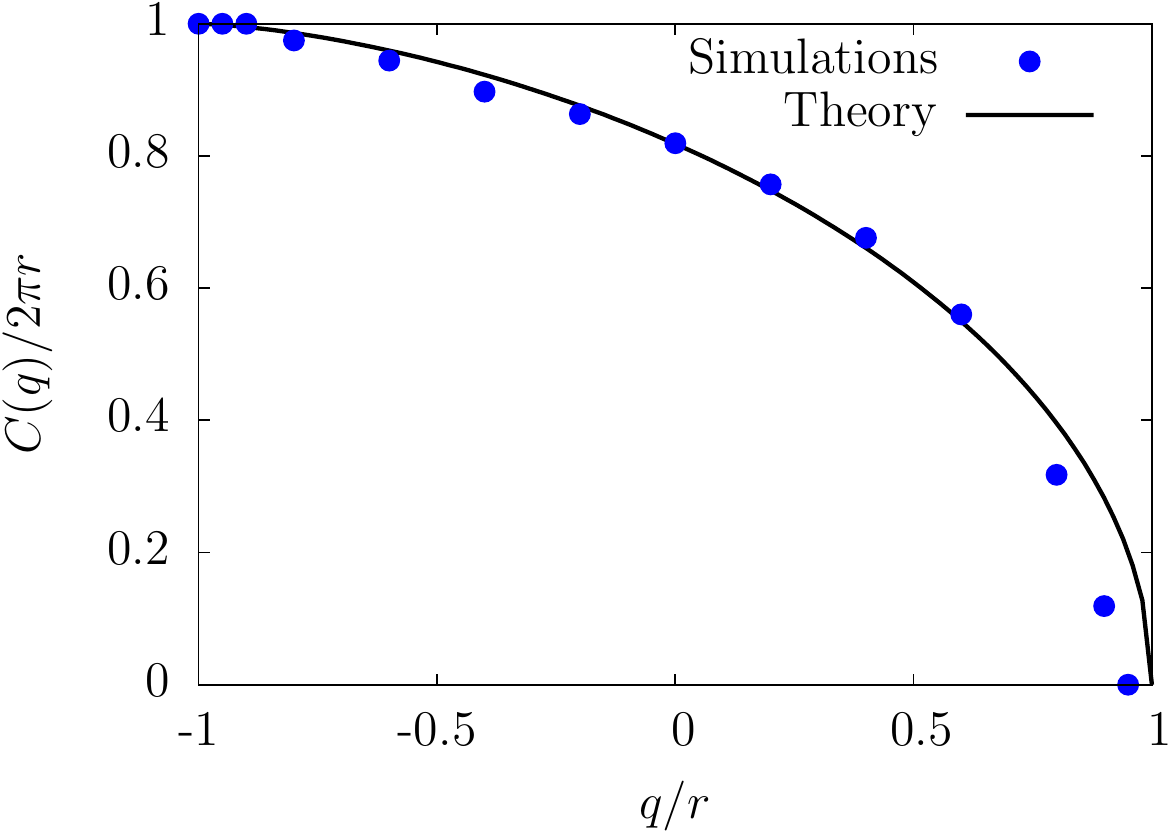}
	\caption{Normalized circumference $C(q)/2\pi r$ against normalized impact factor $q/r$ for a post-reconnection vortex ring. Blue circles indicate numerical data of the mean circumference of the post-reconnection vortex ring.  Solid black curve is the theoretical estimate from Eq.~(\ref{eq:Cpredict}).}
	\label{fig:C_q}
\end{figure}

Figure~\ref{fig:C_q} displays the the numerical data for the post-reconnection vortex ring circumferences and compares to the theoretical prediction given by Eq.~(\ref{eq:Cpredict}). For our setup, we observe very good agreement between the numerical data and the theory.  There is a slight discrepancy at large $q$ which is probably associated with limited resolution.  To further filter out fluctuations, we integrate across the data to get an estimate for the mean circumference of the post-reconnection ring $\left\langle C_{\rm num}\right\rangle= 3.6\times 10^{-1}~\rm cm$ which is remarkably close to the prediction given by Eq.~(\ref{eq:rpredic}): $\left\langle C\right\rangle= 3r/4 = 3.75\times 10^{-1}~\rm cm$.

From the numerical data, we can conclude that on the whole, quantized vortex rings are very robust structures that seem to reconnect in almost an ideal manner.  Moreover, they will preserve a majority ($75\%$) of their vortex line length after reconnection. It stands to assume that vortex rings have the potential to endure multiple sequential vortex reconnections and still transport a substantial fraction of their original energy to the boundaries. Further comment on this point will be made in section~\ref{sec:energytrans}. 

\section{Vortex ring radii in a random tangle}
\label{sec:ringscale}
So far in this article, we have discussed aspects of quantum vortex dynamics involving expected propagation distances and post-reconnection structure.  However, if one wishes to understand the role of vortex rings in quantum turbulence tangles, one must understand the typical scale at which vortex rings are generated. From the previous section, we discussed a  possible mechanism for generating sequentially smaller and smaller vortex rings from an initial size. However, turbulent fluid motion will produce vortex rings at a typical scale associated to the structure of the vortex tangle.  There are many different quantum turbulence tangle structures that can be produced either through superfluid counterflow or large-scale excitations of the flow, both of which will likely affect vortex ring production.   

For isotropic and homogeneous random tangles, Kozik and Svistunov~\cite{kozik_kolmogorov_2008} suggested that the scale at which vortex rings will be created through self-reconnection of the vortex lines will be of the order 
\begin{equation}\label{eq:kozik_radius}
r^*\sim \ell/\left[\ln\left(\ell/a\right)\right]^{1/2},
\end{equation}
where $\ell=L^{-1/2}$ is the inter-vortex scale, and $a$ is the vortex core radius.

We numerically check this hypothesis by creating a spherical vortex tangle created by placing $50$ vortex rings of radius $r=8.0\times 10^{-2}~\rm cm$, randomly distributed and orientated in a sphere of radius $1.0\times 10^{-1}\rm cm$ located in the centre of an open numerical box.  The resulting vortex line density inside the sphere is $L=5.97\times 10^{3}~\rm cm^{-2}$. We evolve the system for a short period of time until the mean curvature is saturated and measure the radial distribution of the emitted vortex rings from the spherical tangle. The fact that the tangle occupies a small volume at the center of the domain leads us to conclude that the measured rings will not have undergone a significant amount of reconnections after being generated.

The radial distribution of the emitted vortex rings is shown in Fig.~\ref{fig:prob_radius}. We observe a very skewed distribution that has a  mean value of $\left\langle r\right\rangle = 1.4\times 10^{-2}~\rm cm$.  However, as indicated by the vertical black dashed line, the estimate of Kozik and Svistunov, $r^*=3.45\times 10^{-3}~\rm cm$, is remarkably close to that of the modal vortex ring radius of $\left\langle r\right\rangle_{\rm mode}=3.34\times 10^{-3}~\rm cm$.  Note that the numerical resolution of this simulation is $2\times 10^{-4}~\rm cm$, an order of magnitude smaller than the most common emitted ring radius.  Moreover, it should be emphasised the significance difference to the radius of initial vortex rings used in constructing the initial tangle. 

\begin{figure}[htbp]
	\centering
	\includegraphics[width = 0.4\textwidth]{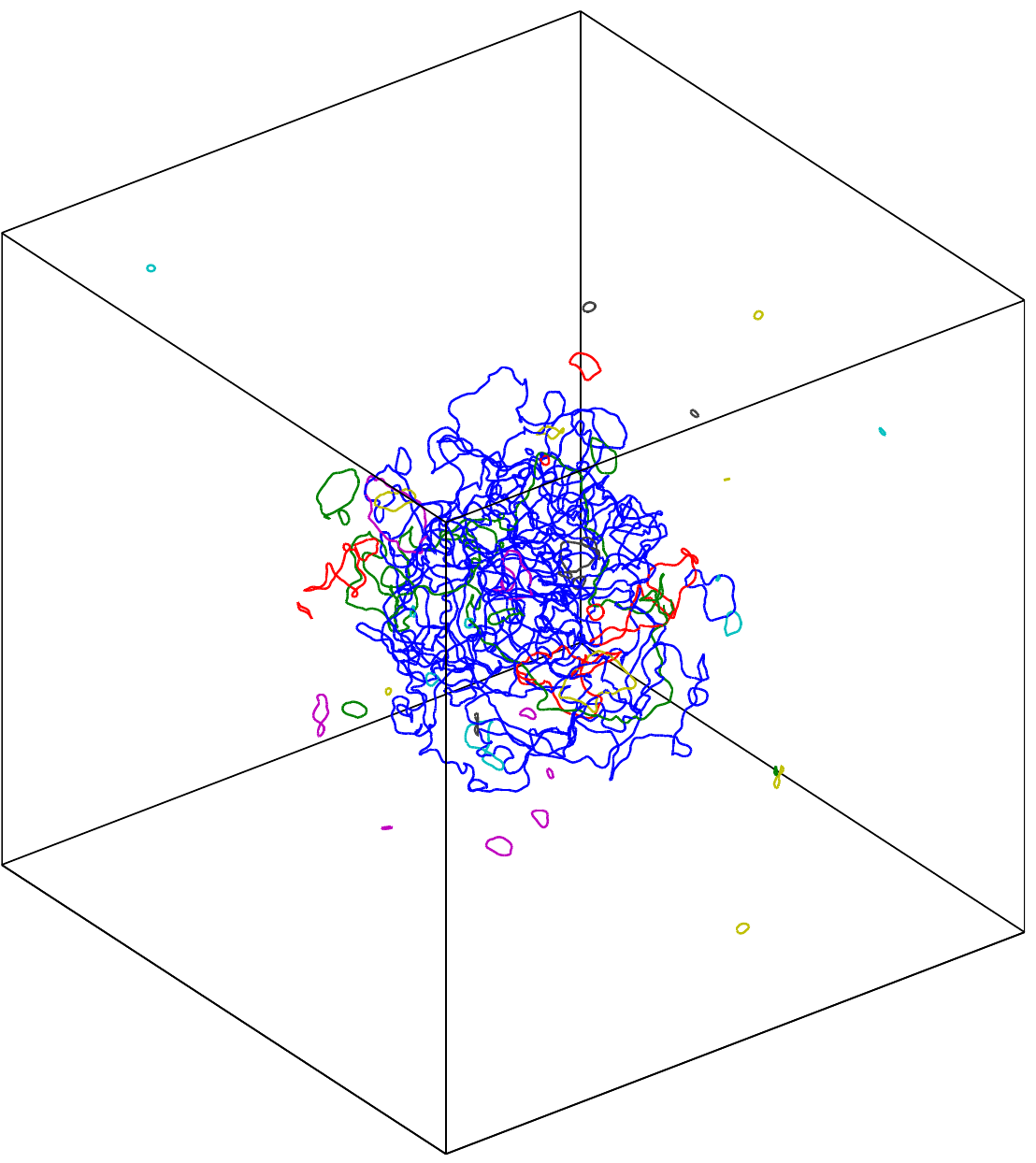}
	\caption{Snapshot of the randomized tangle.  Individual vortex rings are displayed in different colours.}
	\label{fig:coloured_tangle}
\end{figure}

\begin{figure}[htbp]
	\centering
	\includegraphics[width = 0.55\textwidth]{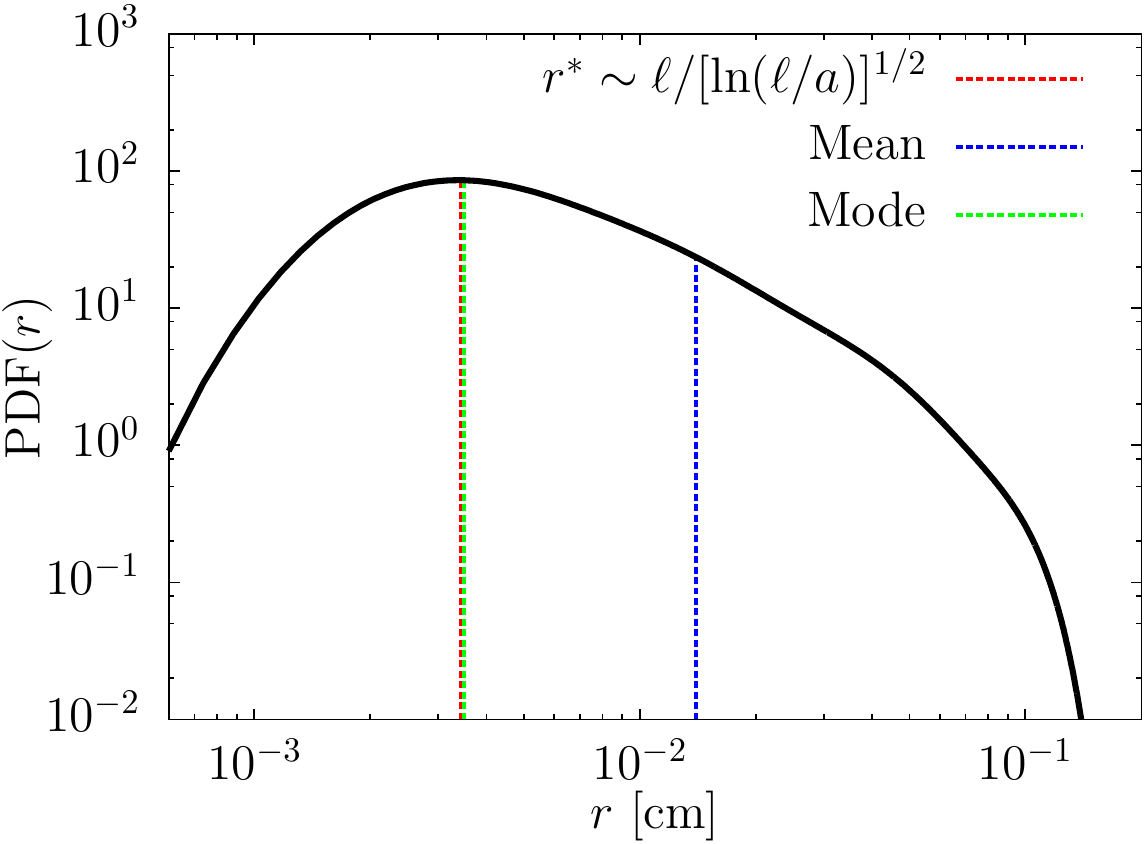}
	\caption{Kernel density estimation of the probability distribution of vortex ring radius in the random tangle presented in Fig.~\ref{fig:coloured_tangle}. The black dashed line indicates Kozik and Svistunov's~\cite{kozik_kolmogorov_2008} predicted radius $r^*\sim \ell/\left[\ln\left(\ell/a\right)\right]^{1/2}=3.45\times 10^{-3}~\rm cm$.}
	\label{fig:prob_radius}
\end{figure}

The numerical evidence seems to concur with the theoretical estimate of Eq.~(\ref{eq:kozik_radius}).  Thus one can imagine that quantum turbulence will create vortex rings roughly at a scale given by~(\ref{eq:kozik_radius}), which then go and propagate further inside the tangle.  It is reasonable to assume that if the tangle is large and dense enough, these vortex rings will likely collide with other filaments and transfer energy and line length throughout the tangle.  In the following section, we try and give a quantitative estimate for this for a typical experimental apparatus.

\section{Energy transport or energy dissipation?}
\label{sec:energytrans}

Given that the energy of a vortex ring is proportional to its radius $E\propto r$, we can attempt to use the numerical results discussed so far to estimate the potential energy transfer to an experimental boundary by a vortex ring. Using formula~(\ref{eq:MFP}) as a measure of the expected distance a vortex ring will travel before undergoing a reconnection and assuming that upon reconnection the radius of the vortex ring shrinks, on average by a factor of $3/4$ [Eq.~(\ref{eq:rpredic})] we can expect that after $n$ reconnections, a vortex ring have travelled on average a distance
\begin{equation}\label{eq:ring_dist}
\left\langle d\right\rangle=\frac{1}{2Lr}\sum_{k=0}^n \left(\frac{4}{3}\right)^k.
\end{equation}
Moreover, one may alternatively ask what is the expected number of reconnections a vortex ring will experience when traveling a distance $d$? Using some straightforward mathematics and the previous formula is
\begin{equation}\label{eq:ring_recon}
n(d) = \left\lfloor \frac{\ln \left(1+2Lrd/3 \right)}{\ln\left(4/3\right)}\right\rfloor -1,
\end{equation}
where $\lfloor \cdot \rfloor$ represents the lowest integer part. Indeed, using formulas~(\ref{eq:ring_dist}) and~(\ref{eq:ring_recon}) we can estimate the expected amount of energy to reach a boundary in a typical experimental apparatus for quantum turbulence. As the energy of a vortex ring is proportional to its radius then after $n$ reconnections the vortex ring will contain a factor of $(3/4)^n$ of its original energy.

For simplicity, let us assume that we are in a one-dimensional domain (length $\mathcal{D}$) and that a vortex ring is equally likely to be produced anywhere along the domain with orientated in either direction.  Then we can calculate the expected energy contained in a vortex ring to reach the boundary to be
\begin{equation}
\left\langle E\right\rangle = \frac{E_0}{\mathcal{D}}\int_{0}^{\mathcal{D}/2}\, \left(\frac{3}{4}\right)^{n(x+\mathcal{D}/2)} + \left(\frac{3}{4}\right)^{n(\mathcal{D}/2 -x)} \, {\rm d}x.
\end{equation}
Typical values of a quantum vortex tangle in an experimental configuration, such as those of the Manchester experimental setup~\cite{walmsley_quantum_2008}, where the experimental cell is a cube with sides of length $\mathcal{D}= 4.5~{\rm cm}$ and with a vortex line density of $L\simeq 1\times 10^3~\rm cm$ (appropriate for the ultra-quantum regime). Then using the results of section~\ref{sec:ringscale} by supposing that the typical vortex ring is created with a radius given by the prediction of Eq.~(\ref{eq:kozik_radius}): $r \simeq 2\times 10^{-3}~\rm cm$, we can expect that on average the energy still remaining in the vortex ring when it approaches a wall will be $\left\langle E\right\rangle\simeq 0.48 E_0$, i.e. roughly $50\%$ of the energy of the initial ring.  However if we consider a line density of $L\simeq 1\times 10^6~\rm cm$ (appropriate for the quasi-classical regime) this estimate drops to around $3\%$.  One should bear in mind that these estimates do not take into account a ring being fully absorbed after a reconnection due to more a more complex tangle geometry. Moreover, it is also important to emphasise that it is unclear whether vortex rings lead to energy dissipation at the boundary or if energy is somehow reflected back towards the tangle.

To summarise our views on energy transport or dissipation, it seems clear that in the ultra-quantum regime ring emission is a plausible and potentially significant energy sink. However, at larger vortex line densities, where quasi-classical behaviour (including agreement with Kolmogorov theory) is seen, dissipation due to Kelvin waves remains the most compelling argument. Indeed even ignoring the fact that the majority of energy is absorbed back into the quasi-classical tangle, the works of Svistunov~\cite{svistunov_superfluid_1995}, Kerr~\cite{kerr_vortex_2011} and Kursa~\etal~\cite{kursa_cascade_2011} shows that vortex ring formation through reconnections is strongly dependent on large angle vortex reconnections, where the vortices are close to being anti-parallel. These types of reconnections are particularly dominant in the ultra-quantum (Vinen and Counterflow) tangles as opposed to quasi-Kolmogorov tangles as noted by our recent work~\cite{baggaley_thermally_2012,laurie_reconnection_2015}.

\section{Conclusions}
\label{sec:con}

To conclude we have investigated the propagation of vortex rings through a tangle of quantized vortex lines. Our goal was to assess if the estimate of the mean free path of a quantized vortex ring propagating in a random tangle gives a representative picture of the ring dynamics.  

We have showed that the mean free path estimate for a vortex ring $\langle d \rangle=1/2rL$ underestimates the distance a ring can propagate before reconnecting with the tangle. This is particular noticeable in the Biot-Savart simulations, and must be due to nonlocal interactions between the ring and the tangle. Moreover, we have also shown, through a combination of simple geometrical arguments and high-resolution numerical simulations that even after a reconnection with a straight vortex line, the vortex ring remains a coherent ring, post-reconnection, with approximately $75 \%$ of its original energy. Using this and data from a recent experiment~\cite{walmsley_quantum_2008} we compute estimates indicating that vortex ring emission could lead to approximately half of the vortex ring energy being transported at the boundaries in the ultra-quantum regime.
Similar estimates for the quasi-classical (polarized) tangle shows only around $3\%$ of energy in vortex rings is transported to the boundaries.  It is still unclear how much energy from a quantum tangle is converted into small vortex ring emissions and so would be an ideal subject for future works. However, we expect that this is more likely in ultra-quantum tangles which have been shown to include more anti-parallel reconnections. With this in mind, we expect that for quasi-Kolmogorov tangles dissipation due to high frequency Kelvin waves remains the most appealing hypothesis.

Finally, we note that much of our analysis is based upon a vortex ring reconnection head on with a straight vortex line. Therefore, further analysis is required to understand the dynamics the vortex rings colliding with tilted vortex lines and other vortex rings.   

\bibliography{biblio}

\end{document}